# Affect Intensity Estimation using Multiple Modalities

## Amol S. Patwardhan, and Gerald M. Knapp

Department of Mechanical and Industrial Engineering
Louisiana State University
apatwa3@lsu.edu

## Abstract

One of the challenges in affect recognition is accurate estimation of the emotion intensity level. This research proposes development of an affect intensity estimation model based on a weighted sum of classification confidence levels, displacement of feature points and speed of feature point motion. The parameters of the model were calculated from data captured using multiple modalities such as face, body posture, hand movement and speech. A preliminary study was conducted to compare the accuracy of the model with the annotated intensity levels. An emotion intensity scale ranging from 0 to 1 along the arousal dimension in the emotion space was used. Results indicated speech and hand modality significantly contributed in improving accuracy in emotion intensity estimation using the proposed model.

## Introduction

With the growing developments in human assistive robotics and smart computer interfaces, the ability of computers to accurately assess human emotional state and tailor responses accordingly will become increasingly important. Many of the automated affect identification studies done to date have only considered binary emotion class models (present, not present). There are considerably fewer methods looking at a magnitude estimation of emotion classes. (Beszedes and Culverhouse 2007) performed a comparison between human emotion estimation and SVM based emotion intensity estimation. The emotion intensity was represented using four emotion levels such as level 0 (neutral), level 1, level 2 and level 3. The SVM-based emotion intensity level estimation performed poorly compared to identification done by human respondents. (Destephe et al. 2013) used low, middle, high and exaggerated enacted emotions for the purpose of emotion intensity representation. (Trabelsi and Frasson 2010) used translation of annotated emotion intensities on a digital scale ranging from -1 to 1, with -1 being extreme negative emotion intensity and 1 being extreme positive emotion intensity. (Park et al. 2008) showed that recognition rate could be used to determine emotion intensity by establishing emotion boundaries in a three dimensional affect space. The research also discussed how emotion intensity influenced use of modality for expressing individual emotions. (Jeni et al. 2013) used action units (AU) and three intensity levels such as absent, low, present to quantify emotion intensity estimation.

In (He, Liu, and Xiong 2008), an emotion model to represent emotion intensity, based on emotion transforms and decay mechanism, was developed. The result used a range of 0 to 1 to quantify emotion intensity using fuzzy inference system and the emotion model. A study (Dhall and Goeke, 2012) used neutral, small smile, large smile, small laugh, large laugh, and thrilled as the emotion intensity levels for predicting intensity using Gaussian Process Regression (GPR) and compared the mean average error with models based on Support Vector Regression (SVR) and Kernel Partial Least Square (KPLS). (Liao, Chuang, and Lai 2012) used a score function to optimize the emotion intensity estimation from facial expressions and compared work to other methods such as AdaSVM, AdaBoost, RankBoost and RegRankBoost except for anger. (Saha and Ghosh 2011) used Local Binary Pattern (LBP) as features and neural networks for determining the emotion intensity estimates on a scale of low, medium and high.

(Meng and Bianchi-Berthouze 2013) showed that transition of emotion levels was slow and these dimension levels could be treated as hidden states in Hidden Markov Models (HMM) framework. Emotion was measured in terms of emotion intensity as a function of time and speed of emotional expressivity (Oda and Isono 2008). Results showed that perception of emotions changed with expression speed. (Zhang and Zhou 2008) showed that maximum intensity and ratio of voiced to unvoiced frames proved to be important discriminating feature in speech based emotion recognition. (Peng, QingShan, and

Metaxas 2009) used a ranking obtained from ordinal relationship in temporal domain, to estimate emotion intensity. In (Miranda et al. 2010), 13 Mel Ceptral Co-efficient Frequencies (MFCCs), and prosodic features (pitch, intensity and formants) were extracted for speech based emotion recognition.

This research focused not only on facial expressions but also hand, body and speech modality for extracting emotion specific cues. The objective of this research was to estimate the emotion intensity levels based on classification rates from various modalities, previous intensity level estimation, and displacement of feature points and speed of feature point movement. Each of these components was assigned a co-efficient (weight) based on the training data and the weighted sum of these components was then normalized to 0 to 1 scale on the arousal dimension in the emotion space. The research used SVM classifiers for emotion prediction from multiple modalities. The predictions were then combined at the decision level and assigned a confidence level based on majority voting. Anger was used as the candidate emotion for experiments.

## Method Overview

The experimental setup consisted of an affect recognition system that used an infrared depth sensing device called Microsoft Kinect. The system provided ability to capture data from facial expression, hand movement, body posture and speech. The participants (5 individuals) were asked to enact anger in front of the sensor and the features from various modalities were captured in a text file. Emotion intensity for each act was labeled by one individual using the intensity annotation tool on a scale between 0 and 1. The training data was used for development of an emotion intensity estimation model. The model accuracy was measured against the data obtained from a sixth individual who was the test participant. The design of the tool was partly motivated from the FeelTrace tool available in a study (Cowie et al. 2000).

This section provides details about the feature vectors used in various modalities. The facial feature points were extracted using the face api available in Kinect SDK. The face api returns a total of 121 facial feature points. Only the frontal feature points are considered for this research resulting in 60 feature points. The feature points include locations on the face such as upper, lower, middle eyebrows on both sides, upper, lower and middle eyelids on both sides, upper, lower and middle points on upper lip and upper, lower and middle points on lower lip. The feature points tracked for body modality were middle, left and right shoulder, right and left elbow, right and left wrist, center spine, right and left hip, center hip and head center.

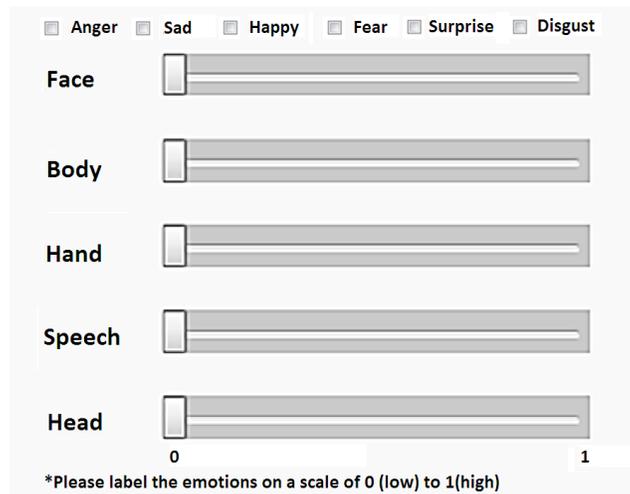

*Figure 1. Emotion Intensity Annotation Tool*

The feature points tracked for hand modality were left and right shoulder joint, left and right elbow and left and right wrist joints. The feature vector definition extracted from various modalities is as follows.

$$FV_m = \{x_0, y_0, x_1, y_1, \ldots, x_n, y_n, \alpha_{ij}, \Delta_i, S_0, \Theta_0, S_1, \Theta_1, \ldots, S_n, \Theta_n\} \quad (1)$$

where m is the modality (face, body, hand), n = 60 for face modality, n = 8 for hand modality, and n = 12 for body modality. i and j $\in$ n. $(x_i, y_i)$ are the co-ordinates of the $i^{th}$ feature point, $\alpha_{ij}$ is the angle between line joining feature point i and feature point j and the horizontal, $\Delta_i$ is the displacement of feature point i across N consecutive frames. $S_i$ is the magnitude component of the velocity of feature point across N consecutive frames. $\Theta_i$ is the orientation component of the velocity of the feature point. The co-ordinates and distance between feature point pairs provided location information about the feature points and velocity of the feature points provided temporal information about the feature points. In case of features from speech modality, this research used openEAR toolkit (Fyben, Wollmer, and Schuller 2009). The baseline feature vector consisting of 988 acoustic features was used. An SVM classifier was trained per modality to recognize anger.

## Emotion Intensity Model

This research proposed an emotion intensity estimation model $I_{im}$ for face, hand and body modality as follows:

$$I_{im} = \Theta_1 C_f + \Theta_2 \Delta d + \Theta_3 S + \Theta_4 \quad (2)$$

where i is the emotion intensity of the current frame, m is the modality, $C_f$ is the confidence level of the recognized emotion, d is the displacement of feature point across N consecutive frames, S is the speed of feature point across N consecutive frames. d was calculated using, the average of displacement of feature points.

$$\Delta d = \sum_{i=0}^{M} \Delta_i / M \quad (3)$$

where i is the feature representing the displacement $\Delta$ in equation 1, 2, and 3 and M is the number of displacement features in a feature vector $FV_m$ s.t m $\in$ {face, body, hand}. Similarly the value for S was the average speed across feature points for a given modality.

$$S = \sum_{i=0}^{M} S_i / M \quad (4)$$

where i is the feature representing the speed $S_i$ in equation 1 and M is the number of Speed features in a feature vector $FV_m$ s.t m $\in$ {face, body, hand}. The entire set of feature points tend to move from a lower to higher intensity or vice versa and hence there is no need to use the displacement and speed for each feature point in the model. The model indicates that the emotion intensity increases with higher displacement and speed of feature points and higher confidence level in the recognized emotion. Thus the emotion intensity estimation model is a linear function of emotion recognition accuracy, speed and displacement of feature points. In case of speech modality the emotion intensity estimation model $I_{is}$ was defined as follows:

$$I_{js} = \sum_{i=1}^{N} \theta_i f v i \quad (5)$$

where j is the current speech feature vector instance, s represents the speech modality, i is the feature extracted using the openEAR toolkit, n = 38 representing the first prosodic features such as intensity and loudness. $\Theta_i$ is the parameter for the speech based intensity estimation model.

For multimodal emotion intensity estimation, a time window T was used for storing the intensities from individual modalities. After the window of time had elapsed, the final multimodal emotion intensity was calculated as follows:

$$I_{mm} = \bar{I}_m \text{ (average of } available \text{ modalities)} \quad (6)$$

where $\bar{I}_m$ represents the average of emotion intensity for *available* modality m. The intensity estimation from all the modalities was not available at the same time. As a result the strategy to use window of time was used to let a collection of intensities to accumulate before the average was calculated leading to a different contribution from each modality.

## Experimental Setup

A total of 5000 frames of enacted anger were captured using the affect recognition system. Feature vector from a series of 10 consecutive frames were annotated with an emotion intensity level between 0 and 1. This training data was used to optimize the cost function defined as follows:

$$CF = \sum_{i=0}^{N} (Y - \hat{I}_t)^2 \quad (7)$$

where Y is the observed emotion intensity and $\hat{I}$ is the estimated emotion intensity. This is a linear regression problem with an objective of finding the optimal values for the parameters $\Theta_i$ that reduces cost function CF.

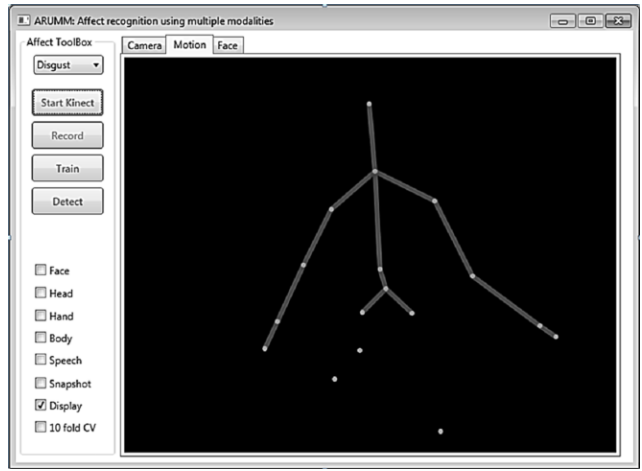

*Figure 2. Feature Point Tracking*

The experiments were performed on the test data which was obtained using 10 fold cross validation. The results of multimodal emotion recognition for anger are shown below:

| Modality | Rate | Precision | Recall |
|---|---|---|---|
| Face | 0.17 | 0.77 | 0.54 |
| Hand | 0.28 | 0.80 | 0.61 |
| Body | 0.21 | 0.79 | 0.43 |
| Speech | 0.24 | 0.68 | 0.41 |
| Multiple | 0.31 | 0.83 | 0.68 |

*Table 1. SVM based Anger recognition results*

Results showed an improvement in the classification rate when multiple modalities were used. The results from hand modality were better compared to other individual modalities due to the wide range of motions providing more discriminative data for detecting anger, compared to other modalities. For the purpose of accuracy calculation, an error margin of +0.1 and -0.1 was used. The emotion intensity estimation based on speech also showed better accuracy compared to the visual modalities. The results of the emotion intensity estimation are shown below:

| Modality | Accuracy |
|---|---|
| Face | 0.37 |
| Hand | 0.55 |
| Body | 0.19 |
| Speech | 0.73 |
| Multiple | 0.46 |

*Table 2. Anger intensity estimation results*

Multimodal emotion intensity estimation showed an improvement in accurate estimation of anger compared to face and body.

## Conclusion

The research showed that emotion intensity can be calculated using a linear estimation model. Although quantitative labeling of emotion intensity is difficult, the strategy of labeling emotion intensity on a scale between 0 and 1 can successfully be used to develop an emotion intensity estimation model. The research also showed that displacement of feature points, speed of feature points can be used for emotion intensity estimation in visual modalities whereas prosodic features can be used for speech modality. The emotion intensity estimation in case of multiple modalities improves the accuracy compared to intensity estimates from individual modalities especially those compared to face and body modalities.